\documentclass[%
aip,
reprint,%
]{revtex4-1}

\usepackage{graphicx}
\usepackage{dcolumn}
\usepackage{bm}

\usepackage[utf8]{inputenc}
\usepackage[T1]{fontenc}
\usepackage{mathptmx}
\usepackage{etoolbox}
\usepackage{color}
\usepackage{siunitx}
\usepackage{amsmath}
\usepackage{amssymb}

\newcommand{\dd}[0]{\mathrm{d}}

\newcommand{\qq}[0]{\boldsymbol{q}}

\newcommand{\PP}[0]{\bm{P}}

\newcommand{\rr}[0]{\bm{r}}

\newcommand{\kB}[0]{k_{\mathrm{B}}}

\newcommand{\uu}[0]{\hat{\bm{u}}}

\newcommand{\MM}[0]{\mathbf{M}}

\definecolor{darkblue}{rgb}{0,0,0.6}
\definecolor{darkred}{rgb}{0.6,0,0}
\definecolor{forestgreen}{rgb}{0.13,0.55,0.13}
\usepackage[colorlinks=true,urlcolor=darkblue,citecolor=darkblue,linkcolor=darkred]{hyperref}

\begin{document}



\title{Dynamic correlations in a polar fluid: confronting stochastic density functional theory to simulations}



\author{Sleeba Varghese}
\affiliation{Sorbonne Universit\'e, CNRS, Laboratoire PHENIX (Physicochimie des Electrolytes et Nanosyst\`emes Interfaciaux), 4 place Jussieu, 75005 Paris, France}

\author{Pierre Illien}
\affiliation{Sorbonne Universit\'e, CNRS, Laboratoire PHENIX (Physicochimie des Electrolytes et Nanosyst\`emes Interfaciaux), 4 place Jussieu, 75005 Paris, France}

\author{Benjamin Rotenberg}
\email[]{benjamin.rotenberg@sorbonne-universite.fr}
\affiliation{Sorbonne Universit\'e, CNRS, Laboratoire PHENIX (Physicochimie des Electrolytes et Nanosyst\`emes Interfaciaux), 4 place Jussieu, 75005 Paris, France}
\affiliation{R\'eseau sur le Stockage Electrochimique de l'Energie (RS2E), FR CNRS 3459, 80039 Amiens Cedex, France}


\date{\today}

\begin{abstract}

Understanding the dynamic behavior of polar fluids is essential for modeling complex systems such as electrolytes and biological media. In this work, we develop and apply a Stochastic Density Functional Theory (SDFT) framework to describe the polarization dynamics in the Stockmayer fluid, a prototypical model of dipolar liquids consisting of Lennard-Jones particles with embedded point dipoles. Starting from the overdamped Langevin dynamics of dipolar particles, we derive analytical expressions for the intermediate scattering functions and dynamic structure factors of the longitudinal and transverse components of the polarization field, within linearized SDFT. To assess the theory’s validity, we compare its predictions with results from Brownian Dynamics simulations of the Stockmayer fluid. We find that SDFT captures the longitudinal polarization fluctuations accurately, while transverse fluctuations are underestimated due to the neglect of dipolar correlations. By incorporating the Kirkwood factor into a modified SDFT, we recover quantitative agreement for both components across a range of dipole strengths. This study highlights the utility of SDFT as a coarse-grained description of polar fluid dynamics and provides insights into the role of collective effects in polarization relaxation.
\end{abstract}


\maketitle 

\section{Introduction}

Understanding the microscopic behavior and organization of polar solvents, such as water, is a fundamental question in the field of chemical physics \cite{Gray1984,Rossky1985,hansen_mcdonald_theory_of_simple_liquids_book}. This challenge has direct relevance to a broad spectrum of applications, ranging from the design of advanced batteries and fuel cells in electrochemistry \cite{Jeanmairet2022} to the behavior of charged soft matter systems such as colloids and biological fluids \cite{Bagchi1991a,Nandi2000,Bagchi2010}. Accurate prediction of solvent structure and dynamics from first principles is especially difficult due to the complex interplay between thermal fluctuations, electrostatic and steric interactions, and the challenge of accounting for the orientation of the molecules.

From a theoretical point of view, two complementary approaches have emerged to address these questions. On the one hand, particle-based simulation methods, such as molecular dynamics, offer detailed insights into the time evolution of individual solvent molecules \cite{Pollock1980, Neumann1983,Hesse-Bezot1984, Neumann1986,Neumann1986a,  bopp_frequency_1998, Bartke2007}. For instance, proposing accurate force fields to describe the structure and dynamics of water has been a central goal of computational physical chemistry during the past decades, resulting in a wealth of well-calibrated options \cite{Stillinger1974, Berendsen1987, Mahoney2000, Rick2004, Abascal2005}.  On the other hand, analytical frameworks enable a more general, and often more computationally efficient, exploration of such systems. Their static structure can be accurately predicted \textit{e.g.} using field-theoretical descriptions \cite{Berthoumieux2018,Berthoumieux2019,Berthoumieux2015a}, or molecular density functional theory \cite{Jeanmairet2013,Jeanmairet2016}. Describing their dynamical response requires other tools, such as dynamical density functional theories or mode-coupling-like theorie. For instance, in a series of works \cite{Bagchi1991a}, Chandra and Bagchi proposed an approach based on dynamical density functional theory (DDFT) \cite{teVrugt2020a}: with appropriate closure approximations, deterministic equations obeyed by the polarization field can be obtained, and their relaxation processes can be characterized \cite{Chandra1988}. Such approaches can be refined by accounting for short-range interactions \cite{Bagchi1989}, and allow to establish relationships between microscopic and collective relaxation times \cite{chandra_relationship_1990}. However, these results were not confronted to microscopic simulations, in such a way that the validity of the approximations at stake are not always clearly outlined.

Alternatively, intrinsically stochastic descriptions can be employed. Within this category, stochastic density functional theory (SDFT) has become an important tool to investigate the dynamics of interacting entities, such as colloids or molecules, on the microscopic and nanoscopic scales \cite{Illien2024d}. Building on the foundational works of Kawasaki \cite{Kawasaki1994} and Dean \cite{Dean1996}, who introduced a formalism to coarse-grain the dynamics of interacting Langevin particles, SDFT provides a powerful theoretical framework to describe the time evolution of density fields under thermal noise. In physical chemistry, this framework was applied to study electrolyte solutions, and more specifically conductivity under various conditions \cite{Demery2016,Peraud2016,Donev2019, Avni2022,Avni2022a,Bernard2023a,Bonneau2023,Hoang2023,Berthoumieux2024}, spatial and temporal ionic correlations \cite{Frusawa2019a,Frusawa2022}, and viscosity in charged fluids \cite{Wada2005,Okamoto2022,Robin2024}. This approach was later extended to include orientational degrees of freedom. In this context, the BBGKY hierarchy of equations and its different closures were discussed \cite{Cugliandolo2015}. Using DDFT-like closure schemes, static as well as dynamic correlations were computed from this approach \cite{Déjardin2018,Déjardin2019,Déjardin2019a}, but were not confronted to numerical simulations, therefore limiting the applicability of the approach to describe real systems.

In this context, we recently proposed a SDFT description of water \cite{Illien2024} as a solvent for ions. In this model, water molecules are represented as dipoles with translational and rotational motion that interact with each other through electrostatic interactions. Importantly, this model does not include any short-range repulsion between the dipoles. They are not meant to represent water molecules: instead, they represent 'polarized blobs', and the parameters of the model can be chosen in such a way that the target properties of water are reproduced. The main outcome of this work are the intermediate scattering functions associated with the longitudinal and transverse polarizations. Derived quantities, such as the power spectral density, are obtained straightforwardly. We showed that this simple representation of the solvent was valid in the limit of low frequencies and small wavevectors.

The goal of the present work is to investigate dynamical correlations in a standard model for dipolar liquids (Stockmayer fluid, consisting of particles interacting via point dipoles and Lennard-Jones potentials), and to assess how well these correlations are captured with SDFT in the linearization approximation and neglecting short-range nonelectrostatic interactions, for which analytical results were obtained in Ref.~\citenum{Illien2024}.

To this end, we first derive analytical expressions for the time-dependent polarization fluctuations from SDFT, building upon our previous framework for polar solvents. We then perform Brownian Dynamics (BD) simulations of the Stockmayer fluid to serve as a microscopic reference. By systematically comparing SDFT predictions with simulation data across various dipole moments and diffusion coefficients, we identify the strengths and limitations of the theory. We find that while SDFT reproduces the longitudinal polarization dynamics quantitatively, it underestimates transverse fluctuations due to the neglect of dipole–dipole correlations. By incorporating the Kirkwood $g$-factor into a modified SDFT formulation, we achieve much improved agreement, particularly for the transverse component. These results establish SDFT as a viable coarse-grained model for capturing the essential features of polarization dynamics in dipolar fluids, and provide a framework for extending this approach to more complex polar media.

\section{Polarization fluctuations in polar fluids} 
\label{sec:Polar}

\subsection{Microscopic description}
\label{sec:Polar:Micro}

The system of interest is made of $N$ solvent molecules in a volume $\mathcal{V}$, represented as dipoles of dipolar moment $p$. Their number density is indicated by $C_s = N/\mathcal{V}$, and their positions and unit orientations at time $t$ are denoted by $\rr_1(t),\dots,\rr_N(t)$ and $\uu_1(t),\dots,\uu_N(t)$ respectively. The microscopic polarization density is defined as:
\begin{equation}
\label{eq:polarization}
	\PP(\rr,t) \equiv \sum_{i=1}^N p \uu_i(t) \delta (\rr-\rr_i(t)).
\end{equation}
In Fourier space, the polarization vector field can be decomposed into a  longitudinal and transverse parts (with respect to the wavevector $\bm{q}$), with Cartesian components
\begin{equation}
\label{eq:defPLPT}
	\tilde P_a(\bm{q},t) = \underbrace{ \frac{q_a q_b}{q^2}  \tilde P_b(\bm{q},t)}_{\equiv \tilde P_{L,a}(\bm{q},t)} + \underbrace{  \left( \delta_{ab} - \frac{q_a q_b}{q^2}  \right)  \tilde P_b(\bm{q},t)}_{\equiv \tilde P_{T,a}(\bm{q},t)} \; ,
\end{equation}
where an implicit sum over index $b$ is assumed. Note that, throughout the paper, we will use the following convention for Fourier transformation: $\tilde{f}(\bm{q}) = \int \dd \bm{r}\ e^{-i \bm{q}\cdot \bm{r}} f(\bm{r})$ and $f (\bm{r}) = \int \frac{\dd \bm{q}}{(2\pi)^3}\  e^{i \bm{q}\cdot \bm{r}} \tilde f(\bm{q})$. The spatio-temporal fluctuations of these components of the polarization are described by the corresponding intermediate scattering functions (ISF),
\begin{equation}
\label{eq:defISF}
    F_{L,T}(q, t) \equiv  \frac{1}{N} \langle \bm{\tilde{P}}_{L,T} (\bm{q},t) \cdot \bm{\tilde{P}}_{L,T} (-\bm{q},0) \rangle \; ,
\end{equation}
where $\langle \dots \rangle$ denotes an ensemble average and we have used the fact that for a bulk fluid the ISF only depends on the magnitude $q$ but not on the orientation. The initial value of the ISF is simply the static polarization structure factor 
\begin{equation}
\label{eq:defSpol}
    S_{L,T}(q) \equiv  \frac{1}{N} \langle | \bm{\tilde{P}}_{L,T} (\bm{q})|^2\rangle \; .
\end{equation}
The static fluctuations of the total dipole $\MM=\tilde{\PP}(\bm{q}=0)=\sum_i p \uu_i$ are often characterized by the so-called Kirkwood $g$-factor:
\begin{equation}
\label{eq:defgK}
    g_K= \frac{\langle \MM^2\rangle}{N p^2}  \; ,
\end{equation}
which is a measure of the local orientational correlations between neighboring dipoles \cite{Kirkwood1939,Fulton1975,madden_consistent_1984}. For uncorrelated dipoles, $\langle \MM^2\rangle=Np^2$ and the Kirkwood factor is equal to 1.

The dynamics is often also considered in the frequency domain, 
introducing the Laplace transforms with (with Laplace variable $s=-i\omega$), $\tilde{\tilde{f}}(\omega)=\int_{0}^{\infty} f(t)\ e^{+i\omega t} dt$,
or the power spectral densities (PSD) also known as dynamic structure factors:
\begin{align}
\label{eq:defPSD}
   S_{L,T}(q,\omega) &= \int_{-\infty}^{\infty} F_{L,T}(q,t)\ e^{+i\omega t} dt 
   = 2 \Re\big[\tilde{\tilde{F}}_{L,T} (q,\omega)\big] \; 
\end{align}
where the last equality follows from the fact $F_{L,T}(\bm{q},t)$ are even functions of time.

\subsection{Properties of interest}
\label{sec:Polar:Observable}

The ISF and PSD, which describe the equilibrium fluctuations of the polarization, are directly related to important quantities describing the linear response of the polar fluid to an external field $\bm{E}_\textrm{ext}(\bm{q},\omega)$. We refer the reader to Ref.~\citenum{Hoang2023} for a more comprehensive discussion of electrical fluctuations in electrolytes and their link between a variety of observable properties, but for the present work we consider the permittivity tensor, which relates the polarization to the local electric field $\bm{P}(\bm{q},\omega)=\varepsilon_0 \left[ \varepsilon(\bm{q},\omega) -\mathbf{I}\right]\cdot\bm{E}(\bm{q},\omega)$, with $\varepsilon_{0}$ the vacuum permittivity. The link between its longitudinal and transverse components and the corresponding susceptibilities depend on the boundary conditions (in particular in simulations using periodic boundary conditions).  For $q\neq0$, one has~\cite{felderhof_fluctuation_1980, pollock_frequency-dependent_1981, caillol_dielectric_1987}
\begin{align}
\chi_L(q,\omega) &= 1-\frac{1}{\epsilon_L(q,\omega)}
\, ,
\label{eq:ChiL}
\\
\chi_T(q,\omega) &= \epsilon_T(q,\omega)-1
\label{eq:ChiT}
\; .
\end{align}
For non-polarizable systems the longitudinal susceptibility is related to the ISF Eq.~\ref{eq:defISF} and its Laplace transform as~\cite{giaquinta_collective_1978, madden_consistent_1984, ladanyi_computer_1999, hansen_mcdonald_theory_of_simple_liquids_book}
\begin{align}
    \chi_{L,T}(q,\omega) &= \frac{\beta N}{V \varepsilon_0 \nu_{L,T}} \left[ S_{L,T}(q) + i\omega \tilde{\tilde{F}}_{L,T}(q,\omega) \right]
    \, ,
    \label{eq:Chi}
\end{align}
where $\beta = 1/k_{B}T$ ($k_{B}$ is the Boltzmann constant and $T$ is the temperature), and where $\nu_L=1$ and $\nu_T=2$. The above results establish a direct link between the microscopic structure and dynamics of the system with key properties of polar fluids. Decades of theoretical approaches with various levels of refinement and computer simulations have already provided many important results. 

 The static permittivity $\varepsilon_r\equiv\varepsilon(q=0,\omega=0)$ of the system described in Section~\ref{sec:Polar:Micro}, which is a material property, is related to the static fluctuations of total dipole (see \textit{e.g.} Ref.~\citenum{madden_consistent_1984}). This link is rather subtle, because the polarization fluctuations also depend on the shape of the considered system and the dielectric properties of the medium in which it is ``embedded''. This is particularly important to compute the permittivity in simulations under periodic boundary conditions. Specifically, for a spherical sample embedded in a medium with permittivity $\varepsilon'$ (see \textit{e.g.} Refs.~\citenum{de_leeuw_simulation_1980,caillol_asymptotic_1992}), one has
\begin{equation}
\label{eq:Kirkwoodembedded}
    \frac{(\varepsilon_r-1)(2\varepsilon'+1)}{2\varepsilon'+\varepsilon_r} = 3 y g_K\; ,
\end{equation}
where $g_K$ defined in Eq.~\eqref{eq:defgK} and $y$ is a dimensionless parameter often found in the literature on polar systems
\begin{equation}
\label{eq:defy}
    y \equiv \frac{C_s p^2}{9\varepsilon_0k_BT} = \frac{C_s\alpha_\textrm{or}}{3} \; .
\end{equation}
In the second expression, we have introduced the orientational polarizability $\alpha_\textrm{or}= p^2 /3\varepsilon_0 k_BT$ characterizing the linear response of a single dipole at temperature $T$ to an external field. This quantity is a characteristic volume and its product with the number density $C_s$ in the $y$ parameter controls the importance of dipolar couplings in the system, just as the product $C_s\Lambda^3$, with $\Lambda$ the de Broglie thermal wavelength, controls that of quantum effects, or the product $C_s \sigma^3$, with $\sigma$ a particle diameter, controls that of excluded volume.

When the permittivity of the embedding medium is the same permittivity as the system of interest ($\varepsilon'=\varepsilon_r$), Eq.~\eqref{eq:Kirkwoodembedded} results in the Kirkwood formula: $(\varepsilon_r-1)(2\varepsilon_r+1)/\varepsilon_r = 9 y g_K$, whereas for $\varepsilon'=1$ one recovers the Clausius-Mosotti equation $(\varepsilon_r-1)/(\varepsilon_r+2)=y g_K$, and in the limit $\varepsilon'\to\infty$ (so-called ``tin-foil'' boundary conditions in Ewald summation used in most simulations), Eq.~\eqref{eq:Kirkwoodembedded} reduces to 
\begin{equation}
\label{eq:Kirkwoodtinfoil}
    \varepsilon_r-1 = 3 y g_K\; .
\end{equation}

Another important physical insight is the anisotropy of the dipole fluctuations. For example, the static longitudinal fluctuations of the dipole are smaller than the transverse ones~\cite{madden_consistent_1984}, by a factor $\epsilon_r$ in the $q\to0$ limit: 
\begin{equation}
\label{eq:staticfluctuationratio}
    \lim_{q\to0} \frac{S_L(q)}{S_T(q)} = \frac{1}{\varepsilon_r}
    \; .
\end{equation}

The dynamics of polarization fluctuations can be characterized by effective relaxation times defined as the integral of the normalized correlation functions,
\begin{equation}
\label{eq:deftau}
    \tau_{L,T}(q) = \int_0^\infty \frac{F_{L,T}(q,t)}{S_{L,T}(q)} \, {\rm d}t = \frac{\tilde{\tilde{F}}_{L,T}(q,0)}{S_{L,T}(q)} \; .
\end{equation}
The longitudinal polarization fluctuations decay faster than the transverse ones, with a ratio of characteristic times~\cite{madden_consistent_1984,Kivelson1989,chandra_relationship_1990} 
\begin{equation}
\label{eq:dynamicfluctuationratio}
    \lim_{q\to0} \frac{\tau_L(q)}{\tau_T(q)} = \frac{1}{\varepsilon_r}
    \; .
\end{equation}
From a microscopic perspective, this can be understood in the limit of a large permittivity: whereas parallel components of the dipoles (i.e. their contribution to the longitudinal quantities) tend to compensate each other and therefore decrease the time it takes to modify $S_L(q=0)$ significantly, their perpendicular components tend to reinforce each other, and therefore make their relaxation slower~\cite{madden_consistent_1984,Kivelson1989}.

Experimentally, one can only measure the $q\to0$ limits of the longitudinal and transverse components of the permittivity tensor, which coincide. It is customary to analyze the permittivity using the Debye model (or a sum of Debye modes):
\begin{align}
    \varepsilon(\omega) &= \lim_{q\to0} \varepsilon_L(q,\omega) = \lim_{q\to0} \varepsilon^\text{SDFT}_T(q,\omega) \nonumber \\
    &= 1 + \frac{\varepsilon_r-1}{1- i \omega \tau_D} \; ,
\label{eq:EpsilonOmega}
\end{align}
where the Debye time $\tau_D$ characterizes the dynamics of collective polarization fluctuations. This time can also be related to the individual dipole dynamics (in the presence of the other dipoles), characterized by: 
\begin{equation}
    \label{eq:taudipole}
    \tau_p = \int_0^\infty \left\langle \hat{\bm{u}}_i(t)\cdot\hat{\bm{u}}_i(0)\right\rangle \, {\rm d}t \; ,
\end{equation}
where the unit vector describing the orientation of dipole $i$ follows Eq.~\eqref{eq:langevin_ua}. Note that this is not the definition in Ref.~\citenum{madden_consistent_1984}, even though it must coincide for exponential relaxation.

\section{Stochastic density functional theory for a polar fluid} 
\label{sec:SDFT}

In this section, we introduce the SDFT description for the dynamics in polar fluids, corresponding to the ion-free case of the more general results presented in Ref.~\citenum{Illien2024} for electrolytes. We then summarize the main results for the polarization fluctuations in the absence of ions, which had not been reported in our previous work. Finally, we discuss the relevance of this level of theory to describe polar fluids, in order to clarify the points to be addressed by comparison with microscopic simulations of such systems.

\subsection{Governing equations}
\label{sec:SDFT:equations}

In the overdamped limit, the evolution equations of the positions and orientations read:
\begin{eqnarray}
	\frac{\dd\bm{r}_i}{\dd t} &=& -\mu_s(p \hat{\bm{u}}_i\cdot \nabla) \nabla\varphi(\bm{r}_i) \nonumber\\
	&&-\mu_s\sum_{j\neq i} \nabla V(\uu_i,\uu_j,\bm{r}_i-\bm{r}_j)  + \sqrt{2D_{s}} \bm{\xi}^{t}_i(t) \label{eq:langevin_rs}\\
	 \frac{\dd \hat{\bm{u}}_i}{\dd t} &=& \left\{-[\mu_s^r p\hat{\bm{u}}_i \times \nabla\varphi(\bm{r}_i)] + \sqrt{2D_s^r} \bm{\xi}_i^{r}(t) \right\}\times \hat{\bm{u}}_i  \label{eq:langevin_ua}
\end{eqnarray}
where $D_{s}$ is the translational diffusion coefficient and $D^{r}_{s}$ is the rotational diffusion coefficient of the polar molecules, with their associated mobilities defined as $\mu_{s} = \beta D_{s}$ and $\mu^{r}_{s} = \beta D^{r}_{s}$. The translational and orientational noises $\boldsymbol{\xi}_i^t(t)$ and $\boldsymbol{\xi}_i^r(t)$ are uncorrelated Gaussian white noises of zero average and unit variance, i.e. $\langle \xi_{i,n}^\alpha (t)\xi_{j,m}^\gamma(t') \rangle = \delta_{\alpha\gamma} \delta_{mn} \delta_{ij} \delta(t-t')$, where $n$ or $m$ are components of the vectors and Greek letters refer to particle labels. 
$V(\bm{r})$ is the interparticle potential, that may account for short-range steric repulsion -- this term is not included in our SDFT analysis, but will be considered in the numerical simulations.   $\varphi(\bm{r})$ is the  electrostatic potential at the position $\bm{r}$ respectively, and is obtained by solving Poisson's equation, 
\begin{equation}\label{eq:Poisson}
		-\nabla^2\varphi(\bm{r})=\frac{\rho_s(\bm{r})}{\varepsilon_0},
\end{equation}
where $\rho_s$ is the charge density associated with the polar molecules.

\subsection{Polarization fluctuations}
\label{sec:SDFT:polarization}

The microscopic polarization density is defined as:
\begin{equation}
	\PP(\rr,t) \equiv \sum_{i=1}^N p \uu_i(t) \delta (\rr-\rr_i(t)).
\end{equation}
The charge density will be considered within the dipolar approximation, so that $\rho_s \simeq -\nabla \cdot \PP$. Incorporating the orientational degrees of freedom to the standard SDFT approach introduced by Dean\cite{Dean1996}, and after linearizing around a constant, isotropic and homogeneous state, we obtain the general equation for the evolution of the  polarization field as\cite{Illien2024} 
\begin{align}
&     \partial_t \bm{P}(\bm{r},t) = D_s \nabla^2 \bm{P} -2D_s^r \bm{P} \nonumber\\
&	+\frac{1}{3}p^2 C_s \nabla(\mu_s \nabla^2 \varphi-2\mu_s^r \varphi) + \bm{\Xi} (\bm{r},t) \;,
\label{eq:lineq_P2}
\end{align}
where $ \bm{\Xi}$ is a Gaussian white noise of zero average and of variance:
\begin{align}
&	\langle \Xi_k(\bm{r},t) \Xi_l(\bm{r}',t') \rangle \nonumber\\
	&= \frac{2 p^2 C_s }{3} \delta_{kl} \delta (t-t') (-D_s \nabla^2+2D_s^r) \delta (\bm{r}-\bm{r}') \; .
\end{align}

Starting from Eq. \eqref{eq:lineq_P2}, we obtain the evolution equations for the longitudinal and transverse components of the polarization field in Fourier space as
\begin{align}
    \label{eq:PP_L}
    \partial_t \bm{\tilde{P}}_L(\bm{q},t) &= 
    - \frac{1}{\tau_{L}(q)} 
    \bm{\tilde{P}}_L +\bm{\Xi}_L(\bm{q},t) \; , \\
	\label{eq:PP_T}
	\partial_t \bm{\tilde{P}}_T(\bm{q},t) &= 
    - \frac{1}{\tau_{T}(q)} 
    \bm{\tilde{P}}_T   +\bm{\Xi}_T(\bm{q},t) \; ,  
\end{align}
where the longitudinal and transverse relaxation rates are
\begin{eqnarray}
    \frac{1}{\tau_{L}(q)} &=& \frac{1+3y}{\tau_s^r}(1+q^2a^2),
        \label{eq:tauL} \\ 
    \frac{1}{\tau_{T}(q)} &=&  \frac{1}{\tau_s^r}(1+q^2a^2).
        \label{eq:tauT}
\end{eqnarray}
The dimensionless parameter $y$ is defined in Eq.~\eqref{eq:defy} and we have introduced the rotational time $\tau_s^r=1/2D_s^r$ for an isolated dipole obeying Eq.~\eqref{eq:langevin_ua}, and a length scale emerging from the ratio of the translational and diffusion coefficients, $a \equiv \sqrt{D_s/2D_s^r}$. The latter defines a crossover between two regimes: for $qa\ll1$, one can consider dipole rotation as sufficiently fast to be averaged, whereas for $qa\gg1$ rotation plays a role in the spatial fluctuations of the polarization. Finally, the noises $\bm{\Xi}_L(\bm{q},t)$ and $\bm{\Xi}_T(\bm{q},t)$ are uncorrelated and have respective variances: 
\begin{multline}
        \langle \Xi_{L,i}(\bm{q},t) \Xi_{L,j}(\bm{q}',t')\rangle = \\
        2  \frac{q_i q_j}{q^2} D_s \kB T \kappa_s(q)^2 \varepsilon_0 (2\pi)^3   \delta(\bm{q} +\bm{q}')\delta(t-t')
\end{multline}
and
\begin{multline}
    \langle \Xi_{T,i}(\bm{q},t) \Xi_{T,j}(\bm{q}',t')\rangle = \\
      2 \left( \delta_{ij} -\frac{q_i q_j}{q^2} \right)D_s \kB T \kappa_s(q)^2 \varepsilon_0 (2\pi)^3   \delta(\bm{q}+\bm{q}')\delta(t-t') 
      \; .
\end{multline}
where we have introduced the reciprocal square of a wavenumber-dependent screening length associated with the polarization charge:
\begin{equation}
    \kappa_S(q)^2 \equiv  \frac{  p^2 C_S}{3\varepsilon_0 \kB T a^2} \left(1+ q^2a^2\right) = \frac{3 y}{a^2} \left(1+ q^2a^2\right).
\end{equation}

From Eqs. \eqref{eq:PP_L}-\eqref{eq:PP_T}, one gets the expressions for correlation functions for the polarization densities in the infinite space limit as 
\begin{align}
    \langle {\bm{\tilde{P}}}_L(\bm{q},t) \cdot \bm{\tilde{P}}_L(\bm{q}',t') \rangle &= \frac{(2\pi)^3 \delta(\bm{q}+\bm{q}') p^{2}C_{s}}{3(1+3y)}  
    e^{-|t-t'|/\tau_{L}(q)} \; , \\
    \langle {\bm{\tilde{P}}}_T(\bm{q},t) \cdot \bm{\tilde{P}}_T(\bm{q}',t') \rangle &= 2  \frac{(2\pi)^3 \delta(\bm{q}+\bm{q}') p^{2}C_{s}}{3} e^{-|t-t'|/\tau_{T}(q)} \; .
\end{align}
For a system with periodic boundary conditions, the corresponding expressions are obtained by changing
\begin{eqnarray}
	(2\pi)^3 \delta(\bm{q}+\bm{q}') \to \mathcal{V} \delta_{\bm{q},-\bm{q}'} 
	\label{eq:FT_pbc}
\end{eqnarray}
and $C_s = N/\mathcal{V}$, resulting in intermediate scattering functions (see Eq.~\eqref{eq:defISF}) decaying as single exponential:
\begin{eqnarray}
    F_{L}(q, t) &=& \frac{p^2}{3} \frac{1}{1+3y} e^{-t/\tau_{L}(q)} \; , \label{eq:ISF_long} \\
    F_{T}(q, t) &=& \frac{2p^2}{3} e^{-t/\tau_{T}(q)} \; .
	\label{eq:ISF_tran}
\end{eqnarray}
Note that the characteristic times $\tau_{L,T}(q)$ introduced in Eqs.~\eqref{eq:tauL} and~\eqref{eq:tauT} are thus consistent with the definition of the effective relaxation times Eq.~\eqref{eq:deftau}. The dynamic structure factors (see Eq.~\eqref{eq:defPSD}) are obtained from Eqs. \eqref{eq:ISF_long} and \eqref{eq:ISF_tran}, resulting in: 
\begin{align}
    S_{L}(q,\omega) &= \frac{p^2}{3} \frac{1}{1+3y} \frac{2\tau_{L}(q)}{1+\omega^{2}\tau_L(q)^{2}} \; , \label{eq:PSD_long} \\
    S_{T}(q,\omega) &= \frac{2p^2}{3} \frac{2\tau_{T}(q)}{1+\omega^{2}\tau_{T}(q)^{2}} \; . 
    \label{eq:PSD_tran}
\end{align}

Finally, we note that the solution of the Poisson equation~\eqref{eq:Poisson} in reciprocal space using Fourier transforms corresponds to Fourier series for a periodic system in the limit of an infinite period. Since we did not consider an additional ``surface'' term (see \textit{e.g.} Refs.~\citenum{de_leeuw_simulation_1980, caillol_asymptotic_1992}), the above results correspond to the tin-foil boundary conditions mentioned in Section~\ref{sec:Polar:Observable}. In particular, the corresponding permittivity is given by Eq.~\eqref{eq:Kirkwoodtinfoil}.

\subsection{Relevance of SDFT for polar fluids}
\label{sec:SDFT:discussion}

As mentioned in Sections~\ref{sec:SDFT:equations} and~\ref{sec:SDFT:polarization}, the above analytical results for the polarization fluctuations were obtained after several approximations. In particular, non-electrostatic interactions are neglected and the dipoles are considered to interact only via the mean-field electrostatic potential induced by the other dipoles. In addition, the coupling between dipoles is considered to be sufficiently small that one can linearize the equations around a constant, isotropic, and homogeneous state. Here, we discuss the relevance of these results for the description of polar fluids, before turning to an explicit comparison with particle-based simulations in Section~\ref{sec:BD}.

We first observe that the SDFT predictions for the ISF (Eqs.~\eqref{eq:ISF_long}-\eqref{eq:ISF_tran}) and PSD (Eqs.~\eqref{eq:PSD_long}-\eqref{eq:PSD_tran}) are consistent with the general results for the $q\to0$ limits of the ratios between longitudinal and transverse static structure factors (Eq.~\eqref{eq:staticfluctuationratio}) and relaxation times (Eq.~\eqref{eq:dynamicfluctuationratio}), provided that the permittivity of the fluid is equal to
\begin{equation}
\label{eq:SDFTpermittivity}
   \varepsilon_r^\text{SDFT}\equiv 1+3y \; . 
\end{equation}
Interestingly, Eqs.~\eqref{eq:ISF_long}-\eqref{eq:ISF_tran} and~\eqref{eq:PSD_long}-\eqref{eq:PSD_tran} show that in the SDFT case, the relations~\eqref{eq:staticfluctuationratio} and~\eqref{eq:dynamicfluctuationratio} also hold for all $q$. This is likely due to the fact that the approximations leading to these results neglect short range correlations and correspond to the hydrodynamic regime $q\to0,\omega\to0$. 
In turn, Eq.~\eqref{eq:Kirkwoodtinfoil} then leads to
\begin{equation}
\label{eq:SDFTKirkwood}
   g_K^\text{SDFT}=1 \; . 
\end{equation}
As noted below its definition, Eq.~\eqref{eq:defgK}, the Kirkwood factor is a measure of the local orientational correlations. The absence of short-range correlations in SDFT is consistent with the mean-field treatment of interactions. 

The SDFT predictions Eqs.~\eqref{eq:ISF_long}-\eqref{eq:PSD_tran} depend on only 4 independent physical parameters: the dipole moment $p$; the dimensionless parameter $y$, or equivalently the number density $C_s$ for a given dipole moment and temperature, characterizing the importance of dipolar couplings; the rotational diffusion coefficient $D_s^r$ corresponding to the rotational time for an isolated dipole; the characteristic length $a$, or equivalently for a given $D_s^r$, the translational diffusion coefficient $D_s$ (see below Eqs.~\eqref{eq:tauL}-\eqref{eq:tauT}).

Introducing Eqs.~\eqref{eq:ISF_long}-\eqref{eq:ISF_tran} in Eq.~\eqref{eq:Chi} provides the longitudinal and transverse susceptibilities. Eqs.~\eqref{eq:ChiL} and~\eqref{eq:ChiT} then yield the corresponding components of the wavelength- and frequency-dependent permittivity tensor. The SDFT prediction for the frequency-dependent permittivity is:
\begin{align}
    \varepsilon^\text{SDFT}(\omega) &= \lim_{q\to0} \varepsilon^\text{SDFT}_L(q,\omega) = \lim_{q\to0} \varepsilon^\text{SDFT}_T(q,\omega) \nonumber \\
    &= 1 + \frac{3y}{1- i \omega \tau_D} \; ,
\label{eq:EpsilonOmegaSDFT}
\end{align}
\textit{i.e.} of the Debye form, with a Debye relaxation time 
\begin{eqnarray}
\label{eq:tauDebyeSDFT}
    \tau_D = \tau_s^r  = \frac{1}{2D_s^r} \; .
\end{eqnarray}
This result is consistent with previous results within the Mean Spherical Approximation~\cite{chandra_relationship_1990}, in which, in contrast with the present approach, short-range interactions between solvent particles are accounted for by setting the pair distribution function to zero below the radius of the particles. This leads to $\tau_D=\tau_T(q\to0)=\epsilon_r \tau_L(q\to0)$, since in the present case $\tau_T(q\to0)=\epsilon_r \tau_L(q\to0)= \tau_s^r$, \textit{i.e.} the rotational time for an isolated dipole. However, the fact that it corresponds to the individual relaxation time for rotation is a consequence of the approximations made to model the polarization fluctuations. Indeed, the dynamics of collective fluctuations are in general not identical to that of individual dipoles, and a dynamic Kirkwood factor $\dot{g}$ can be introduced to link the Debye and individual relaxation times~\cite{madden_consistent_1984}.  We still need to check that SDFT predicts that $\tau_p^\text{SDFT}$, defined by Eq.~\eqref{eq:taudipole}, is equal $\tau_s^r$.

In Ref.~\citenum{Illien2024}, we used the experimental permittivity spectrum of water, which is well described in the low-frequency domain by a single Debye mode with $\tau_D\approx 10$~ps and a relative permittivity $\epsilon_r=78.5$, together with Eqs.~\eqref{eq:EpsilonOmegaSDFT} and~\eqref{eq:tauDebyeSDFT}, to obtain values of $D_s^r=0.05$~ps$^{-1}$ and $y$. Combining $D_s^r$ with the experimental value of the translational (self-)diffusion of water at 25$^\circ$C ($D_s=2.3\,10^{-9}$~m$^2$s$^{-1}$), we estimated the lengthscale $a\approx2.14$~\AA. Finally, assuming a number density equal to that of water at the same temperature, $C_s=55$~M (0.033~\AA$^{-3}$), then enforces the dipole $p=1.4$~D consistent with the value of $y$ determined from the experimental permittivity. While this procedure is consistent with experimental data and results in parameters compatible with physical intuition, a better strategy to assess the range of validity and determine the effective parameters of the coarse-grained theory (SDFT) is to build the latter from a microscopic basis. Therefore, in the following we introduce such a microscopic reference obtained by simulations of a more realistic model of a dipolar fluid.

\section{Brownian Dynamics simulations}
\label{sec:BD}

\subsection{Stockmayer fluid}
\label{sec:BD:Stockmayer}

In order to obtain `exact' results for the polarization fluctuations of a polar fluid with microscopic dynamics described by the overdamped Langevin Eqs. \eqref{eq:langevin_rs}-\eqref{eq:langevin_ua} without resorting to the approximations used in SDFT, it is possible to integrate them numerically in Brownian Dynamics simulations. A standard model for polar fluids is the Stockmayer fluid, consisting of point dipoles $ p \bm{\hat{u}}_i$, where $\bm{\hat{u}}_i$ is the unitary orientation vector of molecule $i$ and $p$ the dipole moment magnitude of the particle, also interacting via Lennard-Jones (LJ) potentials to account for short-range range repulsion as well as dispersion between particles. The pairwise Stockmayer potential between  particles $i$ and $j$ separated by a distance $r$ is thus
\begin{equation}
\label{eq:stockmayer}
V(\uu_i,\uu_j,r) = U_{\text{LJ}}(r) + U_{\text{dipole}}(\uu_i,\uu_j,r) \,
\end{equation}
where
\begin{eqnarray}
U_{\textrm{LJ}}(r) =
\begin{cases}
4 \epsilon\Big[\Big(\frac{\sigma}{r}\Big)^{12}-\Big(\frac{\sigma}{r}\Big)^{6}\Big] 
& \textrm{if} \ r \leq r_c, \\
0 & \textrm{if} \ r > r_c,
\end{cases} \label{eq:lj}
\end{eqnarray}
with $\sigma$ and $\epsilon$ are the LJ distance and energy, respectively, and $r_c$ is a cut-off radius, and
\begin{eqnarray}
\label{eq:dipole}
    U_{\textrm{dipole}}(\uu_i,\uu_j,r) =
     \frac{p^2}{4\pi\varepsilon_{0}}\bigg[\frac{1}{r^{3}}(\uu_{i}\cdot \uu_{j}) 
    -\frac{3}{r^{5}}(\uu_{i}\cdot \bm{r})(\uu_{j}\cdot \bm{r})\bigg] \; .
\end{eqnarray}

In MD simulations of the Stockmayer fluid\cite{Pollock1980, Neumann1983,Hesse-Bezot1984, Neumann1986,Neumann1986a,  bopp_frequency_1998, Bartke2007}, short-range repulsion between particles results in momentum transfer by collision and momentum conservation results in hydrodynamic effects such as long-time tails in the velocity autocorrelation functions~\cite{Harp1970}. In BD simulations, momentum is not conserved because of the energy exchange with the thermal bath, so that such effects may not be captured. The overdamped dynamics of Eqs.~\eqref{eq:langevin_rs} and~\eqref{eq:langevin_ua} do not describe short-time features arising from the inertia of particles, but rather introduce the effect of such high-frequency collective modes on the dynamics at intermediate and long times via the friction and stochastic forces acting on each particle.

From the trajectory of Stockmayer particles, we measure the longitudinal and transverse components of the polarization 
\begin{equation}
\label{eq:Pofq}
	\bm{\tilde{P}}(\qq,t) = \sum_{i=1}^N p \uu_i(t) e^{-i \bm{q}\cdot\rr_i(t)} 
\end{equation}
for selected wavevectors $\qq$ compatible with the periodic boundary conditions, from which we compute the corresponding ISF Eqs.~\eqref{eq:ISF_long}-\eqref{eq:ISF_tran} and PSD Eqs.~\eqref{eq:PSD_long}-\eqref{eq:PSD_tran}. In order to assess the effects of these physical parameters, we consider several systems with different dipole moments $p$ and rotational/translational diffusion coefficients $D_s^r$ and $D_s$, keeping the number density $C_s$ and temperature $T$ constant.

\subsection{Simulation Details}
\label{sec:BD:SimDetails}

Table~\ref{tab:sim_para} summarizes the simulation parameters common to all considered systems. In all case, we consider a fixed temperature $T=298$~K and number density $C_s=0.033$~\AA$^{-3}$ (experimental value for water at this temperature and a pressure of 1~atm), which corresponds to a reduced density $C_s^* = C_s \sigma^3 \approx 0.9$ for the present choice of LJ diameter. The LJ interaction parameters for the Stockmayer fluid were taken from Ramirez \textit{et al.}~\cite{ramirez2002density}. As a reference system, we consider particles with a dipole moment $p = 1.85$~D (or in reduced units $p^{*} \equiv p\sqrt{\beta/\varepsilon_{0}\sigma^{3}} = 6.15$), corresponding to Ref.~\citenum{Jeanmairet2016} (comparing MD simulations of the Stockmayer fluid with the same LJ parameters to molecular Density Functional Theory as a simple model for water), translational diffusion coefficient $D_s=2.3\,10^{-9}$~m$^2$s$^{-1}$ (experimental value for water) and rotational diffusion coefficient $D_s^r=0.05$~ps$^{-1}$ (deduced from the experimental Debye relaxation time of water in Ref.~\citenum{Illien2024}). This reference system corresponds to a dipolar coupling parameter (see Eq.~\ref{eq:defy}) $y=3.84$. The corresponding permittivity $\varepsilon_r\approx140$ (see Ref.~\citenum{Jeanmairet2016}) is much larger than that of water, reflecting the importance of dipolar couplings in this system.  In order to probe a wide range of regimes, we also considered systems with dipole moments in the range $p\in[0.148,1.85]$~D, corresponding to $y\in[0.025,3.84]$.

\begin{table}
    \centering
    \begin{tabular}{c|c|c}
         \textbf{Symbol} & \textbf{Definition} & \textbf{Value} \\
         \hline 
         $\sigma$ & LJ diameter & $3.024\ \textrm{\AA}$ \\
         $\epsilon$ & LJ energy & $0.4412\ \textrm{kcal}\ \textrm{mol}^{-1}$ \\         $T$ & temperature & $298\ \textrm{K}$ \\
         $C_{s}$ & number density & $0.033\ \textrm{\AA}^{-3}$ 
    \end{tabular}
    \caption{Parameters common to all simulations of the Stockmayer fluid. See text for the values of the dipole, rotational and translational diffusion coefficients.}
    \label{tab:sim_para}
\end{table}

Short-range LJ and dipole-dipole interactions are computed using Eq.\eqref{eq:lj} and \eqref{eq:dipole}, respectively, with a real-space cut-off $r_{c} = 7.56$~\AA\ ($\approx 2.5\sigma$). Long-range dipolar interactions are computed using the P3M algorithm~\cite{cerda2008p3m} with a relative root mean square error in the per-atom force calculations below $10^{-4}$. Brownian dynamics (BD) simulations were performed using the large atomic/molecular massively parallel simulator\cite{Thompson2022} (LAMMPS) package, using the \textit{brownian} integrator~\cite{Delong2015, Ilie2015}. Eqs.~\eqref{eq:langevin_rs}-\eqref{eq:langevin_ua} are integrated using a timestep $\delta t = 1$~fs, at a temperature $T = 298$ K. Each simulation consists of 1~ns of equilibration followed by 50~ns of production used for data collection, with observables sampled every 40~fs. As a compromise between computational efficiency and ability to probe the limit of small wave vectors, we use a non-cubic simulation box with dimensions $L_{x}\times L_{y}\times L_{z} = 500 \times 50 \times 50$~\AA$^3$, with periodic boundary conditions in all directions. The smallest accessible wave vector is thus $2\pi/L_{x} \approx 0.0126$~\AA$^{-1}$, which corresponds to $q_\text{min}\sigma \approx 0.039$. The results presented in the following sections are averaged over two independent initial conditions and the uncertainties are reported as the standard error computed across the independent trajectories. See Appendix~\ref{sec:Appendix} for additional computational details on the computation of the dynamic structure factor.

\section{Results and discussion}
\label{sec:Results}

\subsection{Static structure factors and relaxation times}
\label{sec:Results:S_tau}

\begin{figure}
{
    \includegraphics[width=\columnwidth]{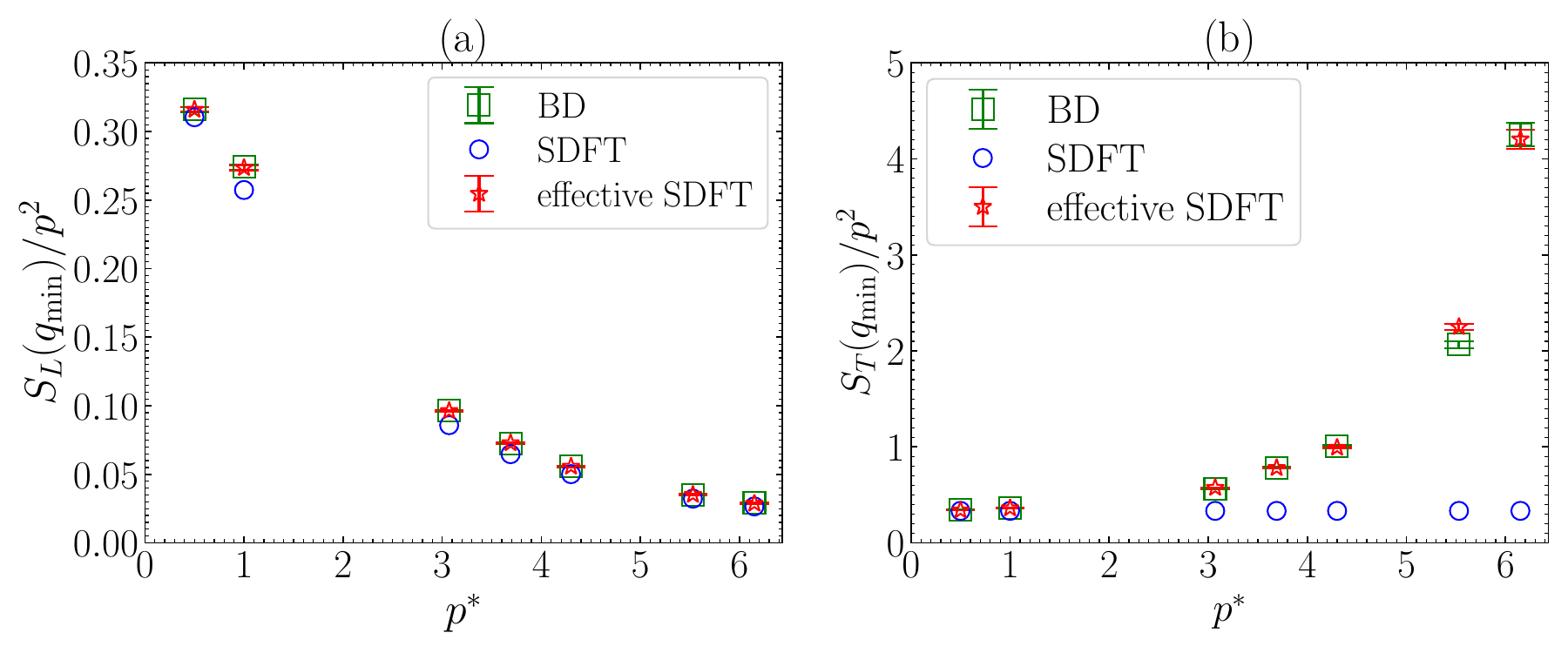}\\
    \includegraphics[width=\columnwidth]{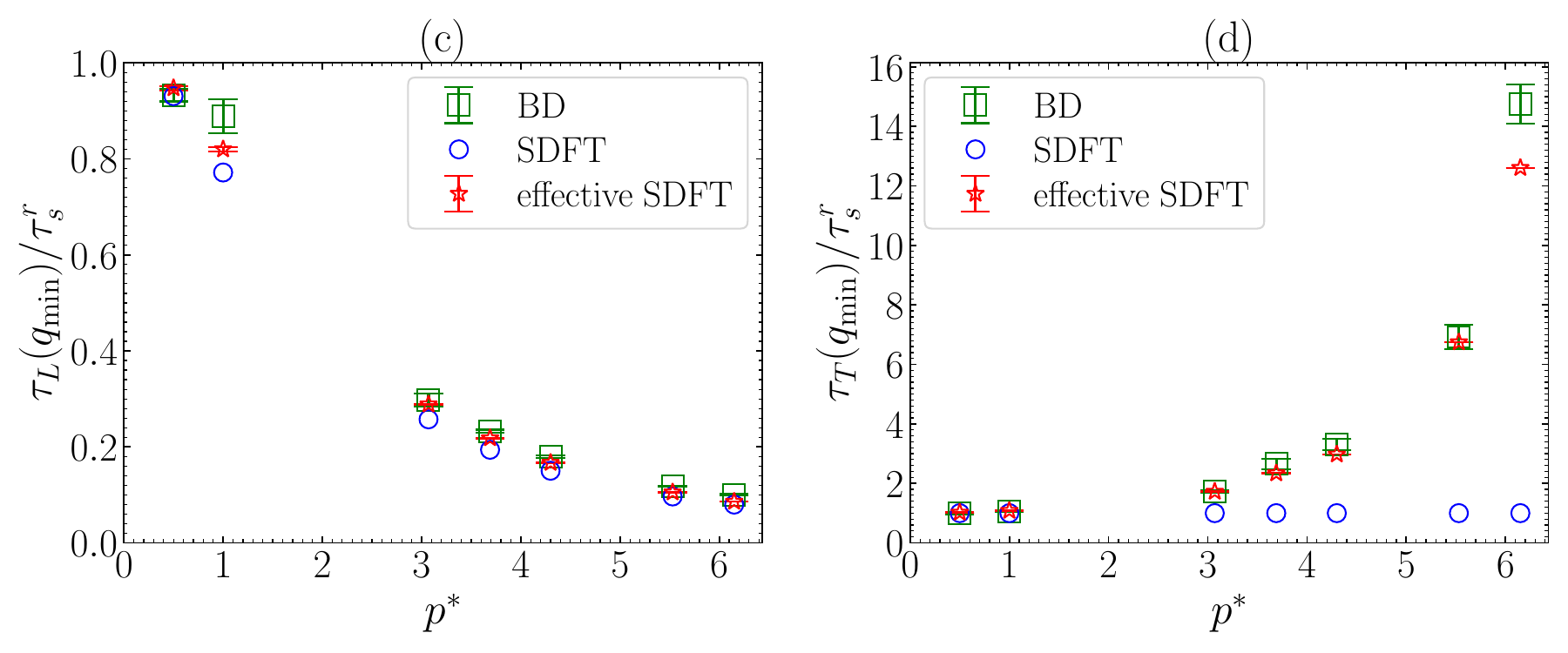}
     \caption{
    Longitudinal (a) and transverse (b) polarization density structure factors and longitudinal (c) and transverse (d) relaxation times, calculated at the smallest considered wavevector $q_\text{min}\sigma \approx 0.039$, as a function of the reduced dipole moment $p^{*} \equiv p\sqrt{\beta/\varepsilon_{0}\sigma^{3}}$. The structure factors and relaxation times are shown normalized by $p^2$ and the rotational diffusion time of a single dipole, $\tau_s^r=1/2D_s^r$, respectively. Green squares indicate the results from Brownian Dynamics simulations of the Stockmayer fluid, while blue circles are the SDFT predictions.  `SDFT' and `effective SDFT'  correspond respectively to the predictions obtained without and with considering the effect of dipolar correlations, respectively. In other words, red stars are the SDFT predictions using effective parameters computed via Eqs.~\eqref{eq:defptilde}, \eqref{eq:defCstilde}, \eqref{eq:defDsrtilde} and \eqref{eq:defDstilde}, using the Kirkwood factor $g_K$ meaured in BD simulations.}
    \label{fig:Sqmin_tauqmin}
    }
\end{figure}

We first compute the longitudinal and static structure factors, whose definition is recalled: $S_{L,T}(q) \equiv  \frac{1}{N} \langle | \bm{\tilde{P}}_{L,T} (\bm{q})|^2\rangle$. Their $q\to 0$ limits (calculated at the smallest considered wavevector $q_\text{min}\sigma \approx 0.039$) are plotted respectively on Figs.~\ref{fig:Sqmin_tauqmin}(a) and~(b), as a function of the reduced dipole moment $p^*$. The BD results for the Stockmayer fluid (green squares) are compared with the prediction from SDFT (\textit{i.e.} $q\to0$ limit of Eqs.~\eqref{eq:ISF_long}-\eqref{eq:ISF_tran}, blue circles), obtained by neglecting non-electrostatic interactions and treating electrostatic couplings at the mean-field level. While SDFT predictions for the $q\to 0$ limit of the longitudinal polarization structure factor, $S_L(q\to0)=p^2/3(1+3y)$, are in excellent agreement with the BD results over the whole dipole range considered, those for the transverse one $S_T(q\to0)=p^2/3$, deviate from the BD results except for small dipoles. 

Identical conclusions can be drawn on the $q\to 0$ limit of the corresponding relaxation times. The latter are determined in BD simulations by fitting the dynamic structure factor $S_{L,T}(q_\text{min},\omega)$ as $S_{L,T}(q_\text{min}) \times 2\tau_{L,T}(q_\text{min})/(1+\omega^2\tau_{L,T}(q_\text{min})^2)$, where the value of the static structure factor, discussed above, is not considered as a fitting parameter (see Appendix~\ref{sec:Appendix} for the actual form taking into account the finite sampling frequency in BD simulations). The BD results for the longitudinal and transverse relaxation times are shown as green symbols in Figs.~\ref{fig:Sqmin_tauqmin}(c) and~(d), together with the corresponding SDFT predictions $\tau_L(q\to0)=\tau_s^r/(1+3y)$ and $\tau_L(q\to0)=\tau_s^r$ (see Eqs.~\eqref{eq:tauL} and~\eqref{eq:tauT}, blue symbols). As for the static structure factor, while SDFT predictions for the $q\to 0$ limit of the longitudinal relaxation time are in excellent agreement with the BD results over the whole dipole range considered, those for the transverse one deviate from the BD results except for small dipoles.

\subsection{Kirkwood factor and permittivity}
\label{sec:Results:gK_epsilon}

\begin{figure}
{  \centering    
    \includegraphics[width=0.95\columnwidth]{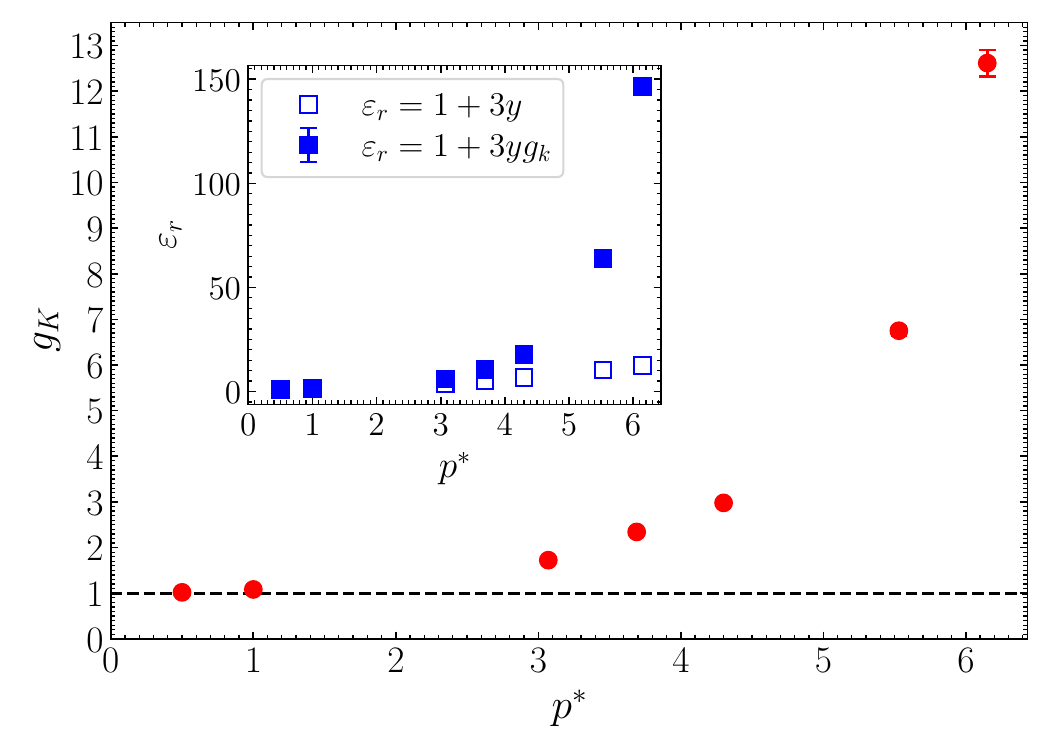} 
     \caption{Kirkwood factor $g_K$ as a function of reduced dipole moment $p^*$, obtained from BD simulations (see Eq.~\eqref{eq:defgK}). The horizontal dashed line corresponds to $g_{K} =1$. The inset shows the corresponding permittivity computed by Eq.~\eqref{eq:Kirkwoodtinfoil} (full symbols), and the prediction assuming $g_{K} =1$ (empty symbols).
    }
    \label{fig:sdft_kirkwood}
    }
\end{figure}

Since the above predictions were obtained by neglecting non-electrostatic interactions and treating electrostatic couplings at the mean-field level, we can anticipate that at least some of their shortcomings are due to the fact that short-range correlations are not properly taken into account. Such correlations between dipole are usually characterized by the Kirkwood factor $g_K$ defined in Eq.~\eqref{eq:defgK} and shown as a function of the reduced dipole $p^*$ in Fig.~\ref{fig:sdft_kirkwood}. Starting from the expected $g_K=1$ (no correlations) in the limit $p^*\to0$, it grows significantly over the considered range of dipoles. As shown in the inset of Fig.~\ref{fig:sdft_kirkwood}, these correlations have a direct impact on the static permittivity $\varepsilon_{r} = 1+3yg_{K}$ (see Eq.~\eqref{eq:Kirkwoodtinfoil}), which rises up to $\epsilon_{r} = 146$, which is close to the value reported in Ref.~\citenum{Jeanmairet2016} for similar simulation parameters as the ones used in this study and much larger than the value predicted by assuming $g_K=1$ (see Eq.~\eqref{eq:SDFTpermittivity}, also shown as dashed line). 
 
Overall, these observations support the claim that the limitations of SDFT to predict the transverse static structure factor and correlation time are related to the neglect of short-range correlations. The effect of such correlations is less pronounced on the longitudinal static structure factor. This can be understood by taking the limit $q\to0$, $\omega\to0$ of Eq. \eqref{eq:Chi}, which yields $\chi_L(q=0,\omega)=\frac{C_s}{\varepsilon_0 \kB T} S_L(q=0)=1-\frac{1}{\varepsilon_r}$, where the second equality originates from Eq. \eqref{eq:ChiL}. Using the relation between $g_K$ and $\varepsilon_r$ [Eq. \eqref{eq:Kirkwoodtinfoil}], and with a similar reasoning for the transverse component, we get
\begin{align}
\label{eq:SLandT}
    S_{L}(q\to0) &= \frac{ S_T(q\to0)}{\varepsilon_r} = \frac{ p^{2} g_{K}}{3\varepsilon_{r}}  = \frac{ p^{2} g_{K}}{3(1+3y g_K)} 
     \; .
\end{align}
While both components are well described by SDFT Eqs.~\eqref{eq:ISF_long}-\eqref{eq:ISF_tran} when $g_K\approx1$ (\textit{i.e.} sufficiently small $p^*$), for $g_K\gg1$ there is a compensation between the numerator and denominator for $S_{L}(q)\approx p^{2}/9y$ (as correctly predicted by SDFT), whereas no such compensation occurs for $S_{T}(q)$. A similar compensation might be at play to rationalize the fact that the longitudinal relaxation time is well reproduced even when dipolar correlations are important, unlike the transverse one. However, to the best of our knowledge, there are no exact relations analogous to Eq.~\ref{eq:SLandT} for the relaxation times (see nevertheless the discussion below of the approximate link between collective and individual dynamics, which would provide a similar argument).

\subsection{Accounting for short-range correlations: effective SDFT}
\label{sec:Results:effective}

It should be no surprise that the approximations leading to the analytical predictions of Section~\ref{sec:SDFT:polarization} fail to capture short-range effects. In fact, one should expect these predictions to be limited to small wavevectors and frequencies. In the following, we show how to introduce the effect of short-range correlations into the above theory via effective quantities (indicated by a tilde), in order to describe polarization fluctuations in the hydrodynamic regime $q\to0,\omega\to0$ without introducing additional complexity to the model.

To that end, we first note that in order to correctly describe the static correlations $S_T(q\to0)$ by the SDFT prediction, which corresponds to $g_K^\text{SDFT}=1$ (see Eq.~\eqref{eq:SDFTKirkwood}), with an effective dipole $\tilde{p}$, Eq.~\eqref{eq:SLandT} implies $\tilde{p}^2 \equiv p^2 g_K$, \textit{i.e.} 
\begin{equation}
    \label{eq:defptilde}
    \tilde{p} \equiv p \sqrt{g_K} = p g_K^{1/2}
\end{equation}
where $g_K$ is the actual Kirkwood factor of the Stockmayer fluid. Previous works have already introduced an effective dipole to account for the effect of short-range correlations on the electrostatic potential or the pair distribution functions at long distance (see \textit{e.g.} Refs.~\cite{Nienhuis1972, Finken2003, Ballenegger2004}), generally leading to an increase of the ``bare'' dipole by a factor $g_K$ (the effective interaction between two distant dipoles being further screened by a factor $\varepsilon_r$). In the present case, the square root can be understood by considering the fact that the description with effective dipoles aims at reproducing collective fluctuations and not the actual structure around a given dipole.  In order to also correctly describe $S_L(q\to0)$, one should also map the permittivity at the microscopic and effective levels of description, \textit{i.e.}
$\varepsilon_r=1+3yg_K=1+3\tilde{y}$, leading to:
\begin{equation}
    \label{eq:defytilde}
    \tilde{y} \equiv y g_K \; ,
\end{equation}
which together with Eq.~\eqref{eq:defptilde} implies
\begin{equation}
\label{eq:defCstilde}
\tilde{C}_s \equiv C_s = C_s g_K^0   \; .
\end{equation}
Kournopoulos \textit{et al.}\cite{Kournopoulos2022} also proposed recently a scheme to describe the static dielectric constant of dipolar fluids from molecular considerations, by introducing a numerical scaling factor in the Kirkwood-Onsager relation to improve the agreement between a mean-field theory and the static permittivity measured in MD simulations of the Stockmayer fluid. In contrast, here we introduce a mapping between two levels of theory and introduce the Kirkwood factor from simulations only for numerical applications.

Turning now to dynamical properties, in order to correctly describe the relaxation times $\tau_{L,T}(q\to0)$ (see Eqs.~\eqref{eq:deftau}) one should consider the effect of short-range correlations on the collective dynamics of dipoles. This issue has been investigated in particular by Kivelson and Madden~\cite{kivelson1975theory, madden_consistent_1984,samanta2022nonlinear}, who showed that by neglecting some dynamical correlations, one obtains a particularly simple relation between the Debye relaxation time  $\tau_D=\tau_T(q\to0)=\epsilon_r \tau_L(q\to0)$ and the correlation time for the reorientation dynamics of a single dipole (in the presence of the others), $\tau_1$, namely $\tau_D= \tau_1 g_K$. The SDFT result Eq.~\eqref{eq:tauDebyeSDFT} suggests that one should rescale the individual rotational time $\tilde{\tau}_s^r\equiv\tau_s^r g_K$, or equivalently the rotational diffusion time:
\begin{equation}
    \label{eq:defDsrtilde}
    \tilde{D}_s^r \equiv D_s^r g_K^{-1}\; .
\end{equation}
Finally, in order to consistently describe the $q-$dependence of the relaxation times with SDFT, one should keep $a$ unchanged (see Eqs.~\eqref{eq:tauL} and~\eqref{eq:tauT}) under rescaling of the rotational diffusion coefficient. This implies the same rescaling of the translational diffusion coefficient, 
\begin{equation}
    \label{eq:defDstilde}
    \tilde{D}_s \equiv D_s g_K^{-1}\; .
\end{equation}

In the following, we will refer to `effective SDFT' when using the SDFT predictions with the effective dipole $\tilde{p}$, concentration $\tilde{C}_s$, rotational diffusion coefficient $\tilde{D}_s^r$ and translational diffusion coefficient $\tilde{D}_s$, obtained by rescaling the corresponding bare quantities with the Kirkwood factor determined from simulations using the appropriate exponents presented in Eqs.~\eqref{eq:defptilde}, \eqref{eq:defCstilde}, \eqref{eq:defDsrtilde} and~\eqref{eq:defDstilde}. To some extent, this effective way to incorporate dipolar correlations through rescaling of the relevant parameters of the model can be seen as analogous to the dressed ion theory for electrolytes, which consists in a self-consistent renormalization of the charge of ions in order to simplify the analytical description of their static correlations~\cite{Kjellander1994,Kjellander2016}.

The predictions of this effective DFT for the static structure factors and relaxation times, also shown in Fig.~\ref{fig:Sqmin_tauqmin}, are in excellent agreement with the BD results. This suggests that incorporating the short-range correlations into SDFT by appropriately rescaling the relevant parameters is sufficient to capture these key observable properties in the hydrodynamic regime $q\to0,\omega\to0$. Before turning to a more stringent test of effective SDFT with the dynamic structure factors in Section~\ref{sec:Results:S_q_omega}, we first consider the static structure factor beyond the $q\to0$ limit.

\subsection{\textit{q}-dependence of the static structure factors}
\label{sec:Results:q-dependence}

\begin{figure}
    \centering       
    \includegraphics[width=\columnwidth]{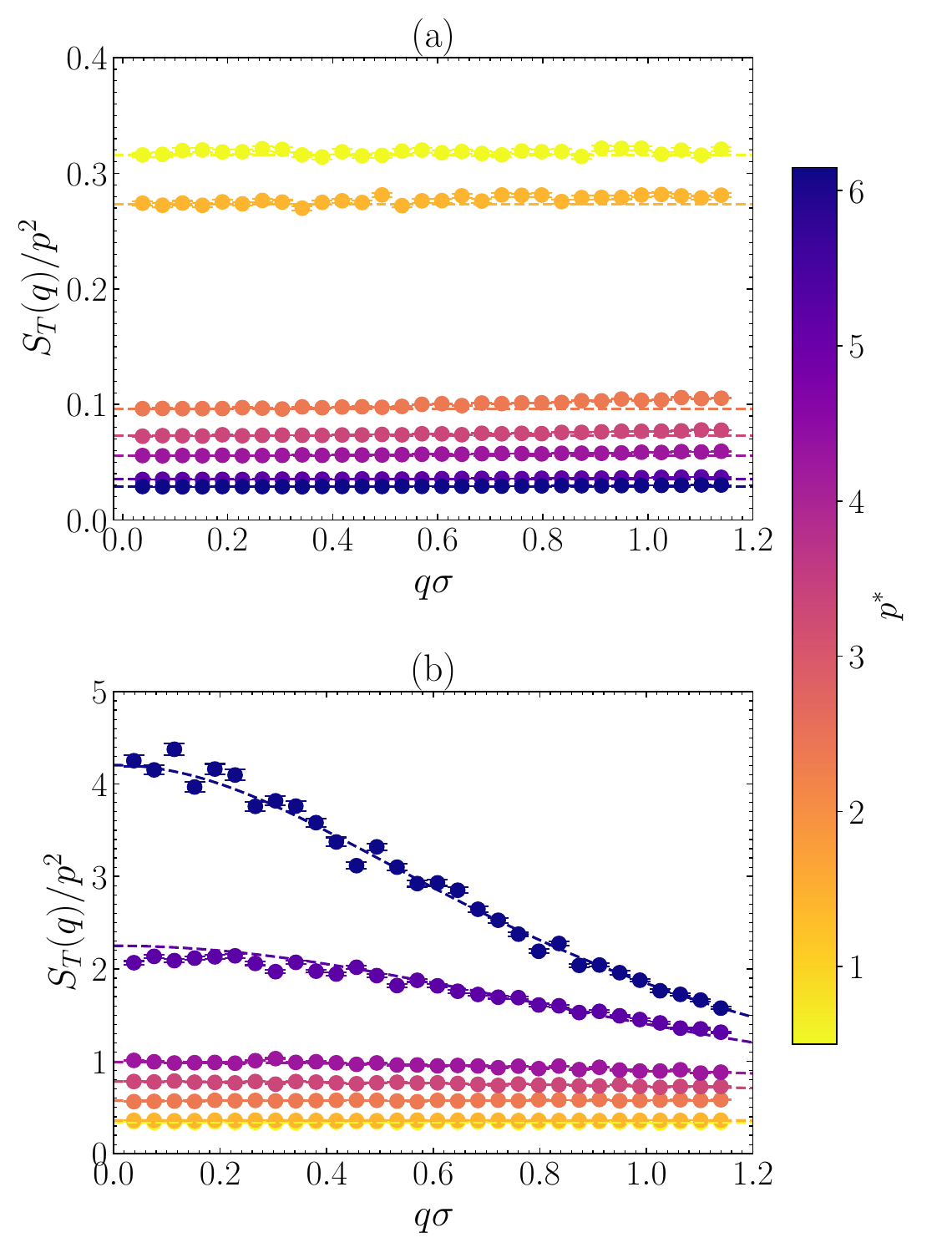} \\
    \caption{Longitudinal (a) and transverse (b) polarization structure factors (see Eq.~\eqref{eq:defSpol}), normalized by $p^2$ and as a function of reduced wave number $q\sigma$, for Stockmayer fluids with reduced density $C_s^*=C_s\sigma^3=0.9$ and various reduced dipoles $p^{*} \equiv p\sqrt{\beta/\varepsilon_{0}\sigma^{3}}$.
    The dashed lines in panels (a) and (b) correspond to $S_{L}(q)/p^2 = g_{K}/3\varepsilon_{r}$ and $S_T(q)/p^2=g_{K}/3(1+q^2\xi^2_{T})$, with $g_K$ and $\varepsilon_r$ from Fig.~\ref{fig:sdft_kirkwood}, and $\xi_T$ a correlation length used as a fitting parameter. 
    Colors encode dipole moments in the range $p^{*}\in[0.5,6.15]$ as indicated by the colorbar.
    \label{fig:sdft_p_Cs}
    }
\end{figure}

The SDFT results obtained by neglecting short-range correlations and treating electrostatic interactions at the mean-field level predict that the static polarization structure factors (corresponding to $t=0$ in Eqs.~\eqref{eq:ISF_long} and ~\eqref{eq:ISF_tran}) are independent of the wavevector $q$. Fig.~\ref{fig:sdft_p_Cs} shows that the BD results for a wide range of dipoles generally satisfy this prediction for wave vectors $q\sigma\lesssim1$, \textit{i.e.} down to wavelengths comparable to the particle size. In addition, the magnitude of the two components in the $q\to0$ limit satisfy the exact results Eq.~\eqref{eq:SLandT}, with the Kirkwood factors $g_K$ and permittivity $\varepsilon_r$ reported in Fig.~\ref{fig:sdft_kirkwood}. 

However, we note that for the largest considered dipoles, the transverse components does depend on the wave vector, in contrast with the simple SDFT prediction. The BD results in that case are well described by $S_T(q)=p ^2g_{K}/3(1+q^2\xi^2_{T})$ where $\xi_T$ is a correlation length used as a fitting parameter. The resulting values are $\xi_T\approx 0.603 \sigma$ and $\xi_T\approx 1.28 \sigma$ for $p^*=5.53$ and 6.15, respectively. These values are comparable to the particle size, consistently with Kirkwood's point that the dipolar correlations remain relatively short-range (see \textit{e.g.} Refs.~\citenum{Kirkwood1939,madden_consistent_1984,zhang2016}). In order to capture these correlations, which are also reflected in the Kirkwood factor, one should go beyond the assumptions leading to the simple analytical results of the present work, namely the dipolar approximation and the omission of any short-range steric repulsion in the SDFT description.

\subsection{Dynamic polarization structure factors}
\label{sec:Results:S_q_omega}

\begin{figure}
    \centering    
    \includegraphics[width=1.0\columnwidth]{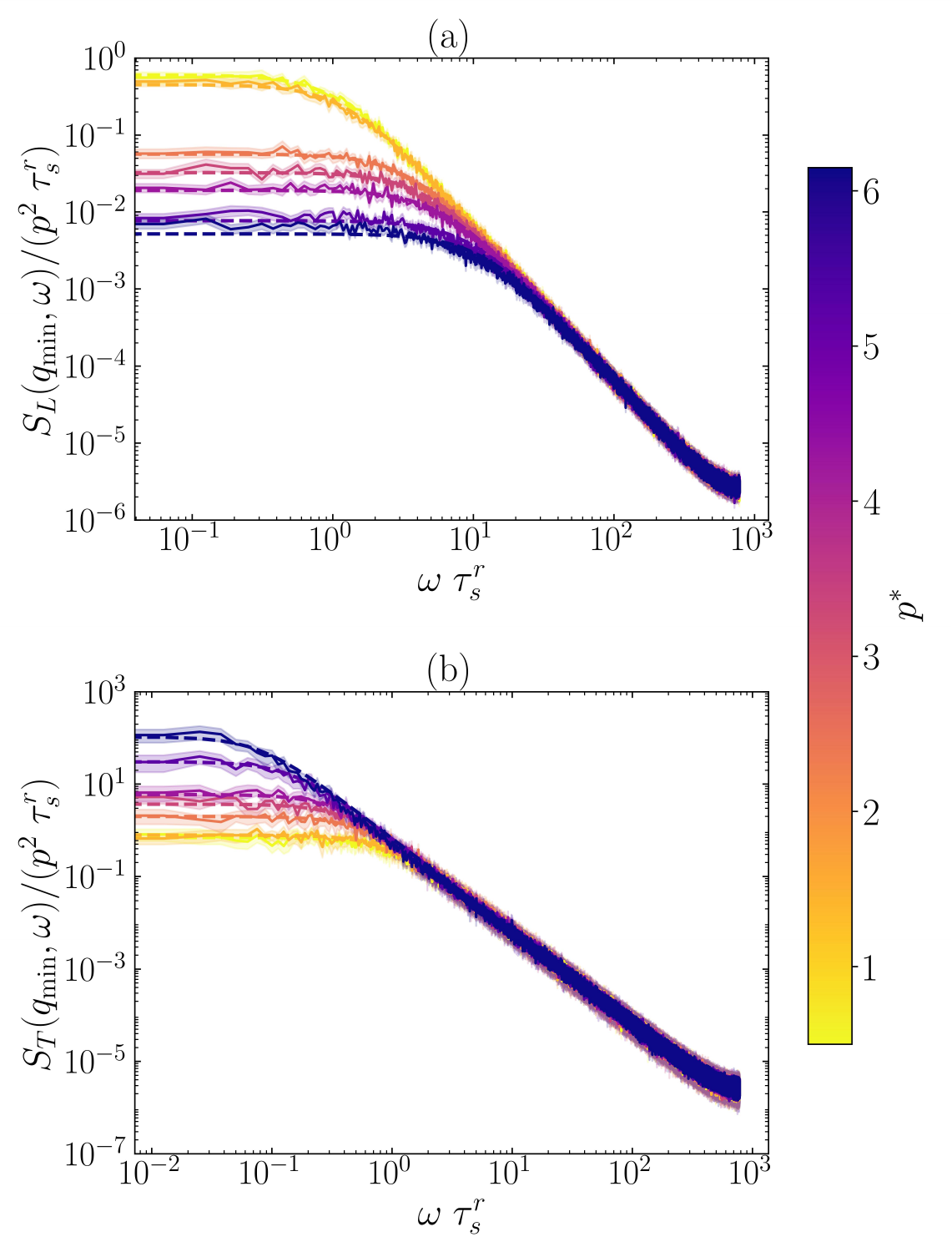} 
     \caption{
     Longitudinal (a) and transverse (b) dynamic polarization structure factors (see Eq.~\eqref{eq:defPSD}), calculated at the smallest considered wave vector $q_\text{min}\sigma \approx 0.039$, for Stockmayer fluids with reduced density $C_s^*=C_s\sigma^3=0.9$ and various reduced dipoles $p^{*} \equiv p\sqrt{\beta/\varepsilon_{0}\sigma^{3}}$ indicated by the color bar. 
     Results are normalized by $p^2\tau_s^r$ and expressed as a function of reduced frequency $\omega \tau_s^r$. The shaded regions denote the uncertainty estimates corresponding to one standard error for the BD results. Dashed lines indicate the effective SDFT predictions, \textit{i.e.}  Eqs.~\eqref{eq:PSD_long}-\eqref{eq:PSD_tran} with effective parameters (see Eqs.~\eqref{eq:defptilde}, \eqref{eq:defCstilde}, \eqref{eq:defDsrtilde} and~\eqref{eq:defDstilde}). 
    \label{fig:S_qmin_omega}
    }
\end{figure}

We finally turn to the polarization dynamics, quantified by the dynamic structure factors $S_{L,T}(q,\omega)$. The BD simulation results for the smallest considered wave vector $q_\text{min}\sigma \approx 0.039$ are shown in Fig.~\ref{fig:S_qmin_omega}. They are very well described by the effect SDFT predictions, \textit{i.e.}  Eqs.~\eqref{eq:PSD_long}-\eqref{eq:PSD_tran} with effective parameters (see Eqs.~\eqref{eq:defptilde}, \eqref{eq:defCstilde}, \eqref{eq:defDsrtilde} and~\eqref{eq:defDstilde}), also shown as dashed lines, over the whole range of considered frequencies. The plateau deviating from the $\omega^{-2}$ scaling at the largest frequencies is due to the finite sampling frequency in simulations and can be captured analytically by considering the discrete Fourier transform (see Appendix~\ref{sec:Appendix}) of the ISF Eqs.~\eqref{eq:ISF_long} and ~\eqref{eq:ISF_tran}. Overall, the results of Fig.~\ref{fig:S_qmin_omega} show that $S_{L,T}(q\to0,\omega)$ for the overdamped Stockmayer fluid correspond to exponentially decaying intermediate scattering functions of the polarization at all times, as predicted by SDFT. Furthermore, they confirm that the simple rescaling of the physical parameters by the Kirkwood factor with appropriate exponents is sufficient to capture the effect of short-range correlations on the dynamical polarization fluctuations. The SDFT approach therefore extends well beyond the static limit, and correctly captures the behavior of the dynamics of the system for a large range of frequencies, when the correlations between dipoles are accounted for in an effective way.

\begin{figure}
        \centering
         \includegraphics[width=\columnwidth]{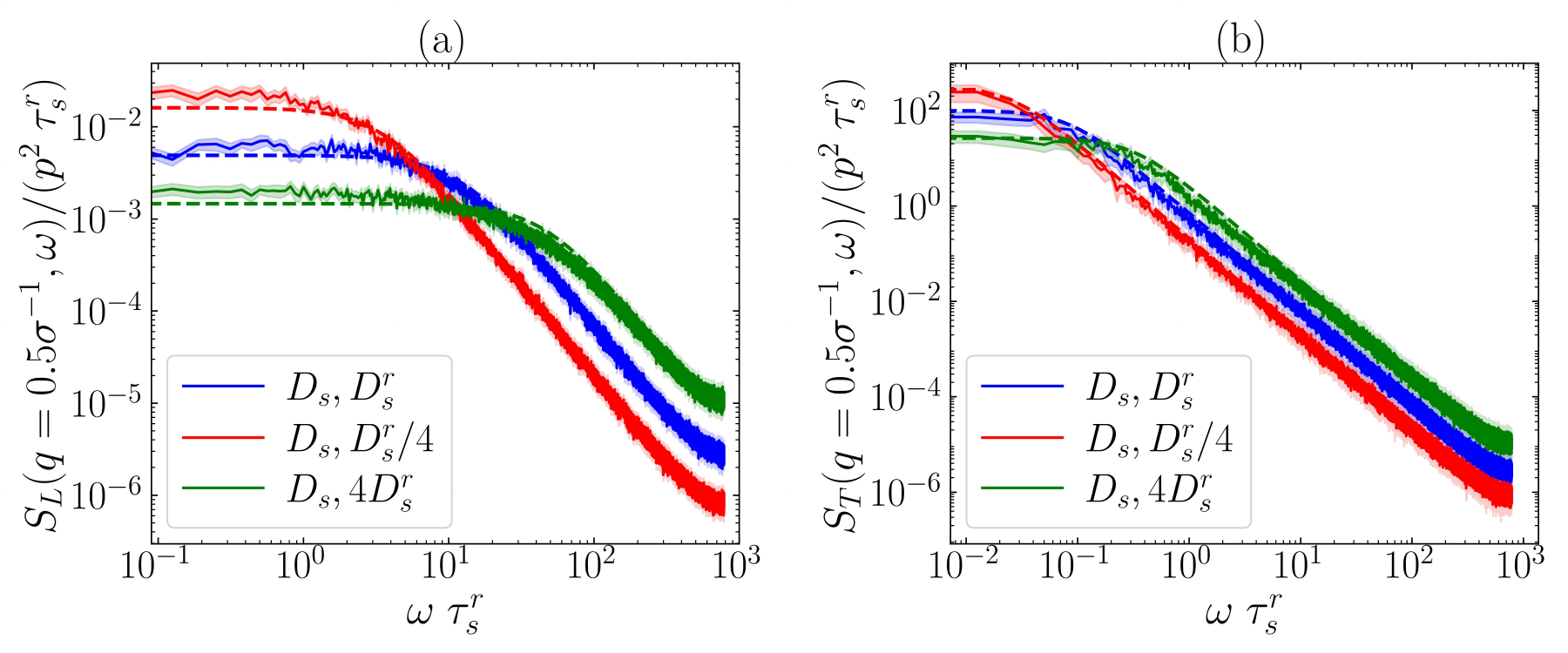}\\ 
         \includegraphics[width=\columnwidth]{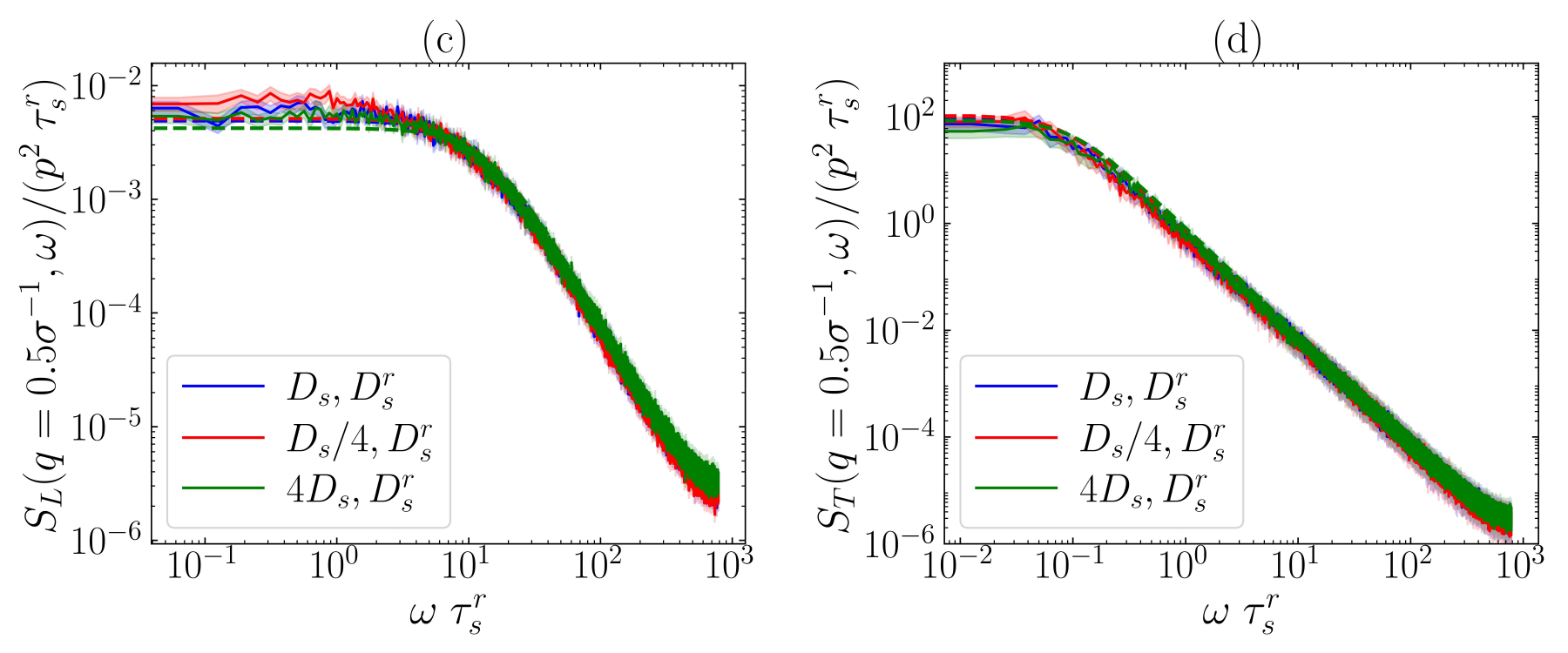} 
        \caption{
        Effect of the translational and rotational diffusion coefficients on the longitudinal [(a) and (c)] and transverse [(b) and (d)] dynamic polarization structure factors (see Eq.~\eqref{eq:defPSD}), calculated at $q_\text{min}\sigma \approx 0.5$, for the Stockmayer fluid with reduced density $C_s^*=C_s\sigma^3=0.9$ and reduced dipoles $p^{*}=6.15$. Starting from a reference system (blue) with $D_s = 2.3 \times 10^{-9}\ \textrm{m}^{2}\textrm{s}^{-1}$, and $D^{r}_{s} = 0.05\ \textrm{ps}^{-1}$, we consider the effect of $D_s^r$ [(a) and (b)]  and $D_s$ [(c) and (d)] by dividing (red) or multiplying (green) them by a factor of 4. The $x$ and $ y$-axes of all plots are normalized by $1/\tau_s^r$ and $p^2\tau_s^r$, respectively, with the rotational diffusion time \textit{of the reference system}. The shaded regions denote the uncertainty estimates corresponding to one standard error for the BD results. Dashed lines indicate the effective SDFT predictions, \textit{i.e.}  Eqs.~\eqref{eq:PSD_long}-\eqref{eq:PSD_tran} with effective parameters (see Eqs.~\eqref{eq:defptilde}, \eqref{eq:defCstilde}, \eqref{eq:defDsrtilde} and~\eqref{eq:defDstilde}). 
        }
        \label{fig:S_q_omega_dsdsr}
\end{figure}

Finally, we investigate the ability of effective SDFT to capture the BD results beyond the $q\to0$ limit, as well as the effect of the translational and rotational diffusion coefficients on the dynamics of the polarization fluctuations. Fig.~\ref{fig:S_q_omega_dsdsr} shows $S_{L,T}(q,\omega)$ for a wave vector $q\sigma =0.5$ corresponding to distances comparable to a few particle diameters, for the Stockmayer fluid with the largest considered dipole ($p = 1.85$~D, \textit{i.e.} $p^{*}=6.15$). In this system, short-range correlations are significant, with a Kirkwood factor $g_{K}\approx12.6$ and a correlation length $\xi_T\approx 1.28\sigma$, so that the static structure factor $S_{T}(q=0.5\sigma^{-1})$ differs significantly from $S_{T}(q\to0)$ (see Fig.~\ref{fig:sdft_p_Cs}). Starting from a reference system with $D_s = 2.3 \times 10^{-9}\ \textrm{m}^{2}\textrm{s}^{-1}$ and $D^{r}_{s} = 0.05\ \textrm{ps}^{-1}$, we consider the effect of $D_s^r$ and $D_s$ by multiplying or dividing them by a factor of 4. The BD results show that, for the considered wave vector, the effect of $D^{r}_{s}$ is significant, while that of $D_s$ is very limited, as expected in the small $q$ regime (in the reference case, $a\approx 0.5\sigma$, and in the modified ones $a\approx 0.25\sigma$ or $\sigma$, leading to $qa<1$ in all cases). Out of this regime, $D_s$ might also play a role on the polarization dynamics~\cite{Chandra1988}. Increasing $D_s^r$ results in a decrease of the characteristic time (larger crossover frequency) and of the plateau at low frequency (note that the frequency and dynamic structure factors in Fig.~\ref{fig:S_q_omega_dsdsr} are adimensionalized using $\tau_s^r$ of the reference system). Importantly, Fig.~\ref{fig:S_q_omega_dsdsr} shows that the BD results and the effects of $D_s^r$ and $D_s$ are well captured by effective SDFT even for this case beyond the $q\to0$ limit of Fig.~\ref{fig:S_qmin_omega}.

\section{Conclusion and perspectives}

In this work, we developed and validated a stochastic density functional theory (SDFT) framework to describe the polarization dynamics of polar fluids, using the Stockmayer fluid as a prototypical model. Analytical expressions were derived for the intermediate scattering functions and dynamic structure factors of longitudinal and transverse polarization components under linearized SDFT assumptions. These predictions were then systematically compared with Brownian Dynamics simulations, which served as a microscopic reference.

Our results demonstrate that SDFT, which neglects short-range correlations and treats dipole–dipole interactions at the mean-field level in its current form, provides accurate predictions for the longitudinal polarization fluctuations. However, it underestimates the amplitude and relaxation time of transverse fluctuations, especially for fluids with strong dipolar interactions. We show that this discrepancy arises from the neglect of local orientational correlations, which can be effectively captured by introducing the Kirkwood factor into a modified SDFT. This effective theory, based on parameter rescaling guided by simulation data, achieves quantitative agreement with both static and dynamic properties in the long-wavelength limit.

Beyond validating the theoretical framework, this work provides practical guidelines for the use of SDFT to model real polar solvents and their role in complex environments. In particular, it sets the stage for investigating fluctuation-induced phenomena in electrolytes,\cite{Mahdisoltani2021a,Mahdisoltani2021c,Du2024,Du2025} where polar solvent dynamics could play a critical role, and has been overlooked so far. Furthermore, the model can be extended to account for spatial confinement—an essential factor in many nanoscale systems, including ionic channels, membranes, and porous materials. Capturing how confinement alters polarization correlations and collective relaxation could significantly enhance our understanding of nanoscale transport processes. By providing a consistent and computationally tractable approach, SDFT thus offers a promising path toward bridging molecular-scale interactions with mesoscopic behavior in a wide range of soft and electrochemical systems.


\section*{Acknowledgments}
This project received funding from the European Research Council under the European Union’s Horizon 2020 research and innovation program (grant agreement no. 863473).  P.I. acknowledges financial support from Agence Nationale de la Recherche through project TraNonEq (ANR-24-CE30-0651).

\section*{Author declarations}

\subsection*{Conflict of interest}
There is no conflict of interest to declare.

\subsection*{Author contributions}

\textbf{Sleeba Varghese:} Conceptualization (supporting); Formal analysis (equal); Investigation (lead); Methodology (supporting); Writing/Original Draft Preparation (lead); Writing – review \& editing (equal).
\textbf{Pierre Illien:} Conceptualization (equal); Formal analysis (equal); Investigation (supporting); Methodology (equal); Supervision (supporting); Writing – review \& editing (equal).
\textbf{Benjamin Rotenberg:} Conceptualization (equal); Formal analysis (equal); Funding Acquisition (lead); Investigation (supporting); Methodology (equal); Supervision (lead); Writing – review \& editing (equal).

\section*{Data availability}

\textcolor{red}{To be completed consistently with the final version of the manuscript:} The original data presented in this study are openly available in Zenodo at \textcolor{red}{[DOI/URL]} or \textcolor{red}{[reference/accession number]}.

\appendix
\section{Additional computational details}
\label{sec:Appendix}
The partial Fourier components of the longitudinal and transverse polarization densities, $\bm{\tilde{P}}_{L,T}(\bm{q},t)$, are sampled every 40~fs during the production run of the simulation, for selected wave vectors in the range $q\sigma \in [0.039,11]$, satisfying $q = m 2\pi / L_x$ with $m \in \mathbb{N}$ to ensure compatibility with periodic boundary conditions. The dynamic polarization structure factor is computed for each wave vector using the Fast Fourier Transform algorithm, applied over $n$ blocks of the time series assumed to be statistically independent ($n=50$ corresponding to 1~ns for the longitudinal polarization density, $n=10$ corresponding to 5~ns for the transverse one). The results are further averaged over the two independent initial conditions, with uncertainties estimated as the standard error across the independent trajectories.

To incorporate the effect of finite sampling frequency into the analytical predictions, we express the dynamic polarization structure factors using the discrete Fourier transform of Eqs.~\eqref{eq:ISF_long}-\eqref{eq:ISF_tran}. Hence, Eqs.~\eqref{eq:PSD_long}-\eqref{eq:PSD_tran} become
\begin{equation}
    \label{eq:PSD_DFT}
    S_{L,T}(q, \omega_{k}) = \frac{2 S_{L,T}(q)}{\tau_{L,T} (q)} \bigg|\frac{1}{1-r(\omega_{k})} \bigg|^{2} (\Delta t)^{2} \; ,
\end{equation}
where for each discrete frequency $\omega_{k}= 2\pi k / N_b\Delta t$,
with $\Delta t =40 \delta t$ the interval between successive samples, $N_{\mathrm{b}}$ the total number of samples in a block and $k \in \{0,1,2,...,N_{\mathrm{b}}\}$, we introduced
\begin{equation}
    r(\omega_{k}) = \mathrm{exp}\bigg[-\bigg( \frac{\Delta t}{\tau_{L,T}} + \mathrm{i} \omega_k \Delta t \bigg)\bigg] \; .
\end{equation}

\bibliography{biblio}

\end{document}